\documentclass[manuscript=article]{achemso}            		

\setkeys{acs}{usetitle=true}

\usepackage{graphicx}
\usepackage{subcaption}
\usepackage{multirow}
\usepackage{multicol}
\usepackage[version=3]{mhchem}
\usepackage{longtable}
\usepackage{siunitx}
\usepackage{lmodern}
\usepackage{hyperref}
\usepackage{algorithm}
\usepackage{algpseudocode}
\usepackage{amsmath}
\usepackage{pdfpages}
\DeclareSIUnit\angstrom{\text {Å}}
\usepackage{xr}
\newcommand{\shellcmd}[1]{\texttt{\footnotesize #1}}

\title{Materials Graph Library (MatGL), an open-source graph deep learning library for materials science and chemistry}
\author{Tsz Wai Ko}
\affiliation[UCSD]{Aiiso Yufeng Li Family Department of
Chemical and Nano Engineering, University of California San Diego, 9500 Gilman Dr, Mail Code 0448, La Jolla, CA 92093-0448, United States}
\email{t1ko@ucsd.edu}
\author{Bowen Deng}
\affiliation[UCB]{Department of Materials Science and Engineering, University of California Berkeley, Berkeley, CA, USA}
\alsoaffiliation[MSD]{Materials Sciences Division, Lawrence Berkeley National Laboratory, California 94720, United States}
\author{Marcel Nassar}
\affiliation[Intel]{Intel Labs, Santa Clara, CA, United
States}
\author{Luis Barroso-Luque}
\affiliation[UCB]{Department of Materials Science and Engineering, University of California Berkeley, Berkeley, CA, USA}
\alsoaffiliation[MSD]{Materials Sciences Division, Lawrence Berkeley National Laboratory, California 94720, United States}
\author{Runze Liu}
\affiliation[UCSD]{Aiiso Yufeng Li Family Department of
Chemical and Nano Engineering, University of California San Diego, 9500 Gilman Dr, Mail Code 0448, La Jolla, CA 92093-0448, United States}
\author{Ji Qi}
\affiliation[UCSD]{Aiiso Yufeng Li Family Department of
Chemical and Nano Engineering, University of California San Diego, 9500 Gilman Dr, Mail Code 0448, La Jolla, CA 92093-0448, United States}
\author{Elliott Liu}
\affiliation[UCSD]{Aiiso Yufeng Li Family Department of
Chemical and Nano Engineering, University of California San Diego, 9500 Gilman Dr, Mail Code 0448, La Jolla, CA 92093-0448, United States}
\author{Gerbrand, Ceder}
\affiliation[UCB]{Department of Materials Science and Engineering, University of California Berkeley, Berkeley, CA, USA}
\alsoaffiliation[MSD]{Materials Sciences Division, Lawrence Berkeley National Laboratory, California 94720, United States}
\author{Santiago Miret}
\affiliation[Intel]{Intel Labs, Santa Clara, CA, United
States}
\author{Shyue Ping Ong}
\email{ongsp@ucsd.edu} 
\affiliation[UCSD]{Aiiso Yufeng Li Family Department of
Chemical and Nano Engineering, University of California San Diego, 9500 Gilman Dr, Mail Code 0448, La Jolla, CA 92093-0448, United States}
\date{}

\begin{document}
\maketitle

\begin{abstract}
Graph deep learning models, which incorporate a natural inductive bias for a collection of atoms, are of immense interest in materials science and chemistry.  Here, we introduce the Materials Graph Library (MatGL), an open-source graph deep learning library for materials science and chemistry. Built on top of the popular Deep Graph Library (DGL) and Python Materials Genomics (Pymatgen) packages, our intention is for MatGL to be an extensible ``batteries-included'' library for the development of advanced graph deep learning models for materials property predictions and interatomic potentials. At present, MatGL has efficient implementations for both invariant and equivariant graph deep learning models, including the Materials 3-body Graph Network (M3GNet), MatErials Graph Network (MEGNet), Crystal Hamiltonian Graph Network (CHGNet), TensorNet and SO3Net architectures. MatGL also includes a variety of pre-trained universal interatomic potentials (aka ``foundational materials models (FMM)'') and property prediction models are also included for out-of-box usage, benchmarking and fine-tuning. Finally, MatGL includes support for Pytorch Lightning for rapid training of models.

\end{abstract}
		
\section{Introduction}

In recent years, machine learning (ML) has emerged as a powerful new tool in the materials scientist's toolkit\cite{chenCriticalReviewMachine2020, schmidtRecentAdvancesApplications2019, westermayrPerspectiveIntegratingMachine2021,oviedoInterpretableExplainableMachine2022}. Sophisticated ML models have found their way into a multitude of applications. Surrogate ML models for ``instant'' predictions of properties such as formation energies, band gaps, mechanical properties, etc. \cite{chenGraphNetworksUniversal2019, schmidtCrystalGraphAttention2021, gasteiger2020directional, gasteiger2020fast, pmlr-v139-satorras21a, liu2022spherical,brandstetter2022geometric,kaba2022equivariant,yan2022periodic} have greatly expanded our ability to explore vast chemical spaces for new materials. In addition, Machine learning (ML) has been widely used for parameterizing potential energy surfaces (PESs), enabling the direct prediction of potential energies, forces, and stresses based on atomic positions and chemical species.
These ML interatomic potentials (MLIPs)\cite{schuttSchNetDeepLearning2018a, pmlr-v139-schutt21a, bartokGaussianApproximationPotentials2010a, behlerGeneralizedNeuralNetworkRepresentation2007a, thompsonSpectralNeighborAnalysis2015, drautzAtomicClusterExpansion2019a, batznerE3equivariantGraphNeural2022,koAccurateFourthGenerationMachine2023,kocerNeuralNetworkPotentials2022,koFourthgenerationHighdimensionalNeural2021,koRecentAdvancesOutstanding2023,unkeMachineLearningForce2021a,equiformer} have provided us with the means to parameterize complex PESs to perform large-scale atomistic simulations with unprecedented accuracies. 

\begin{figure}[H]
    \includegraphics[width=\textwidth]{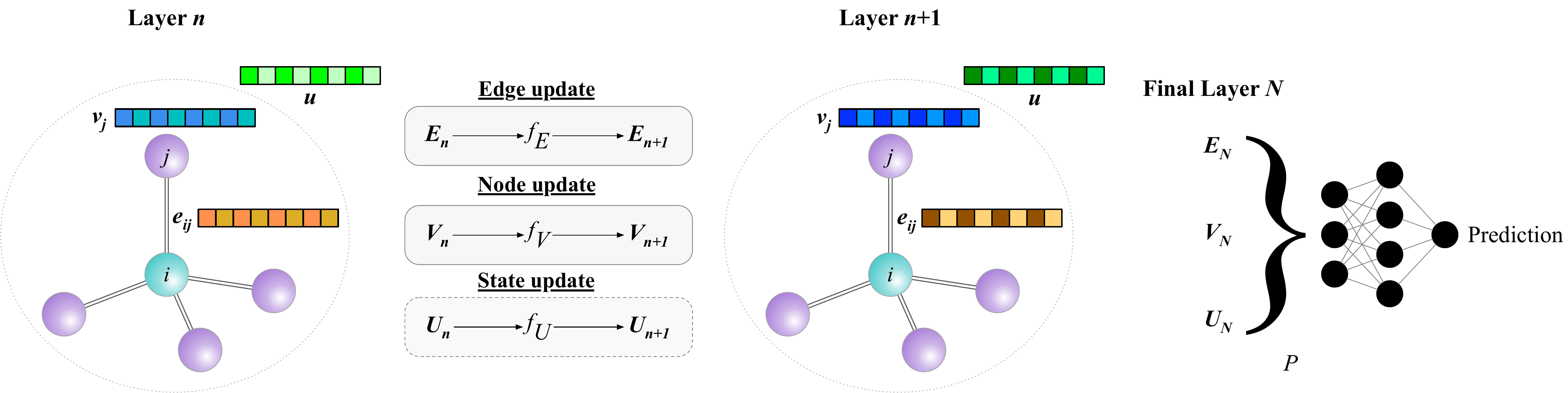}
    \caption{\textbf{Graph deep learning architecture for materials science.} $\mathbf{V_n}$ and $\mathbf{E_n}$ denotes the set of node/atom ($\{\mathbf{v_i}\}$) and edge/bond features ($\{\mathbf{e_{ij}}\}$), respectively, in the $n^{th}$ layer. Some implementations include a global state feature ($\mathbf{U}$) for greater expressive power. Between layers, a sequence of edge ($f_E$), node ($f_V$) and state ($f_U$) update operations are performed. $f_E$, $f_V$ and $f_U$ are usually modeled using multilayer perceptrons. In the final step, the edges, nodes and state features are pooled ($P$) and passed through a multilayer perceptron to arrive at a prediction.}
    \label{fig:graphrepresentation}
\end{figure}

Among ML model architectures, graph deep learning models, also known as graph neural networks (GNNs), utilize a natural representation that incorporates a physically intuitive inductive bias for a collection of atoms.\cite{battaglia2018relational} Figure \ref{fig:graphrepresentation} depicts a typical graph deep learning architecture. In the graph representation, the atoms are nodes and the bonds between atoms (usually defined based on a cutoff radius) are edges. In most implementations, each node is represented by a learned embedding vector for each unique atom type (element). 
Additionally, some architectures such as the MatErials Graph Network (MEGNet)\cite{chenGraphNetworksUniversal2019} and Materials 3-body Graph Network (M3GNet)\cite{chenUniversalGraphDeep2022} also include an optional global state feature ($\mathbf{u}$) to provide greater expressive power, for instance, in the handling of multifidelity data.\cite{chenLearningPropertiesOrdered2021, ko2024data} A graph deep learning model is constructed by performing a sequence of update operations, also known as message passing or graph convolutions. In the final layer, the embeddings are pooled and passed through a final MLP layer to arrive at a final prediction. GNNs can be broadly divided into two classes in terms of how they incorporate symmetry constraints. Invariant GNNs use scalar features such as bond distances and angles to describe the structure, ensuring that the predicted properties remain unchanged with respect to translation, rotation, and permutation. Equivariant GNNs, on the other hand, go one step further by ensuring that the transformation of tensorial properties, such as forces, dipole moments, etc. with respect to rotations are properly handled, thereby allowing the use of directional information extracted from relative bond vectors. 
For a comprehensive overview of different GNN architectures and their applications, readers are referred to recent literature\cite{han2024survey, duval2023hitchhiker}.
Given sufficient training data, GNN architectures such as Nequip\cite{batzner2022}, MACE\cite{batatia2022mace}, Equiformer\cite{liao2022equiformer} and many others\cite{wang2024enhancing, frank2024euclidean, gasteiger2021gemnet} have been shown to provide state-of-the-art accuracies in the prediction of various properties and PESs.\cite{chenGraphNetworksUniversal2019,fungBenchmarkingGraphNeural2021,bandiBenchmarkingMachineLearning2024,fu2022forces} Furthermore, unlike other MLIP architectures based on local-environment descriptors, GNNs have a distinct advantage in the representation of chemically complex systems. The recent emergence of universal MLIPs\cite{chenUniversalGraphDeep2022, dengCHGNetPretrainedUniversal2023,parkScalableParallelAlgorithm2024a,batatia2024foundationmodelatomisticmaterials,barroso2024open,neumann2024orb} (uMLIPs) encompassing the entire periodic table of elements is a particularly effective demonstration of the ability of GNNs to handle diverse chemistries and structures, and such MLIPs can be considered as foundation materials models (FMMs).

At the time of writing, most software implementations of materials GNNs\cite{pelaezTorchMDNet20Fast2024,schuttSchNetPack20Neural2023, axelrodExcitedStateNonadiabatic2022} are for a single architecture, built on PyTorch-Geometric\cite{fey2019fast}, Tensorflow\cite{tensorflow} or JAX\cite{jax2018github}. However, recent benchmarks show that the Deep Graph Library (DGL)\cite{wang2019deep} outperforms PyTorch-Geometric in terms of memory efficiency and speed, particularly when training large graphs under the same GNN architectures for various benchmarks\cite{huangCharacterizingEfficiencyGraph2022,wang2019deep}. This improved efficiency enables the training of models with larger batch sizes as well as the performance of large-size and long-time-scale simulations. 

In this work, we introduce the Materials Graph Library (MatGL), an open-source modular, extensible graph deep learning library for materials science. MatGL is built on DGL, Pytorch and the popular Python Materials Genomics (Pymatgen)\cite{ong2013pymagtgen} and Atomic Simulation Environment (ASE)\cite{larsen2017atomic} materials software libraries. MatGL provides a user-friendly workflow for training property models and MLIPs, with data pipelines and Pytorch Lightning training modules designed for the unique needs of materials science. In its present version, MatGL provides implementations of several state-of-the-art invariant and equivariant GNN architectures, including the Materials 3-body Graph Network (M3GNet)\cite{chenUniversalGraphDeep2022}, MatErials Graph Network (MEGNet)\cite{chenGraphNetworksUniversal2019}, Crystal Hamiltonian Graph Neural Network (CHGNet)\cite{dengCHGNetPretrainedUniversal2023}, TensorNet\cite{simeon2024tensornet} and SO3Net\cite{schuttSchNetPack20Neural2023}, as well as pre-trained FMMs and property models based on these architectures. To facilitate the use of pre-trained FMMs in atomistic simulations, MatGL also implements interfaces to widely used simulation packages such as the Large-scale Atomic/Molecular Massively Parallel Simulator (LAMMPS) and the Atomic Simulation Environment (ASE). The intent for MatGL to serve as a common platform for the scientific community to collaboratively advance graph deep learning architectures and models for materials science.

\section{Results}
In the following sections, we present the MatGL framework, with the manuscript organized as follows: We start with a schematic overview of the core model components, followed by a concise summary of the data pipeline and preprocessing steps. We then introduce the available graph neural network (GNN) architectures for property prediction and the construction of MLIPs. Next, we detail the key components involved in training and deploying these architectures, explaining their integration into MatGL. Additionally, we introduce the simulation interfaces for atomistic simulations and the command-line interface for various applications. Finally, we demonstrate the performance of different GNN architectures on widely used datasets, encompassing both molecular and periodic systems.

\section{Overview}
\begin{figure}[H]
    \centering
    \includegraphics[width=\textwidth]{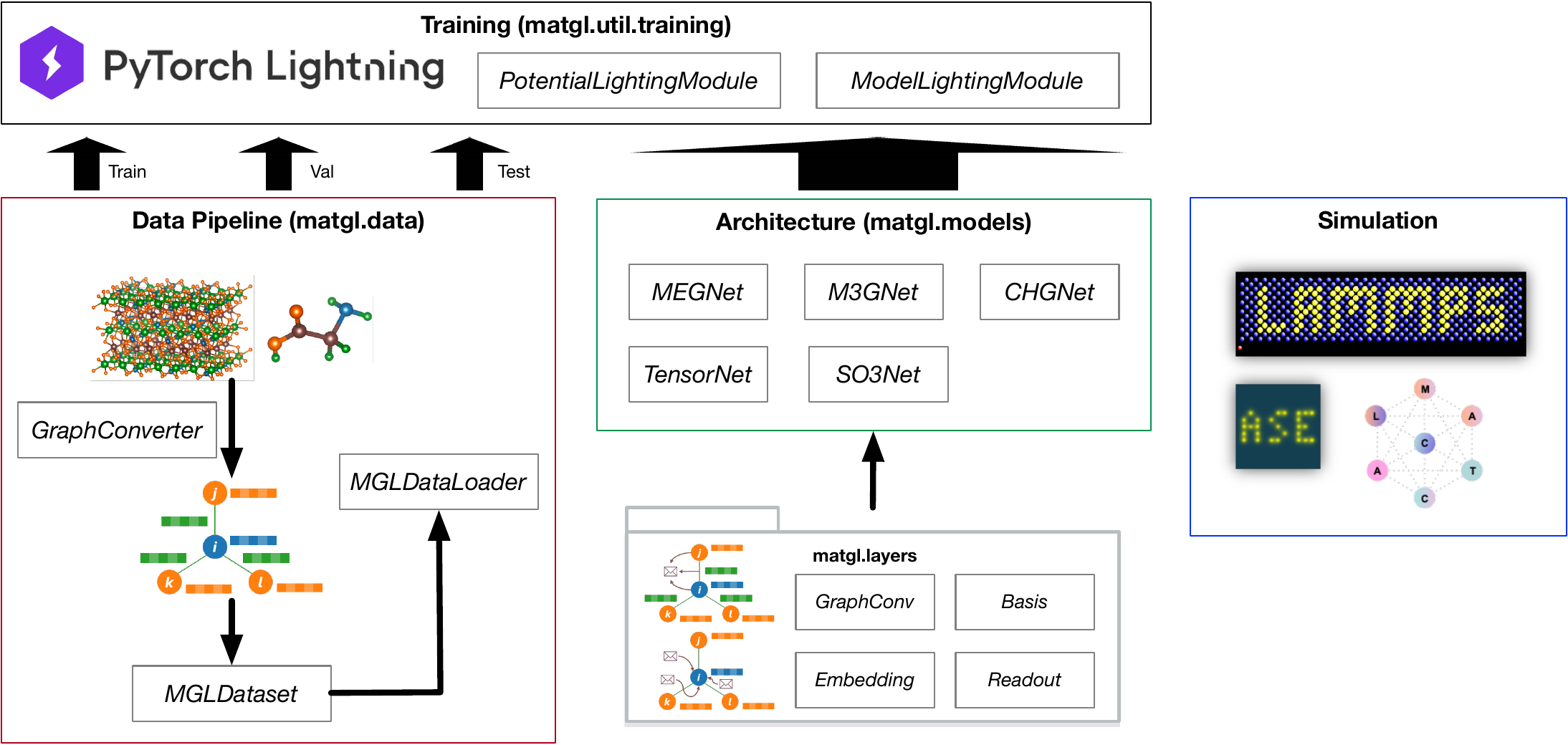}
    \caption{\textbf{Overview of MatGL.} Class names are in italics. MatGL can be broken down into four main components: 1. the data pipeline component preprocesses a set of raw data into graphs and labels; 2. the architecture component build the GNN model using modular layers implemented; 3. the training component utilizesPyTorch-Lightning to train either property models or MLIPs; and 4. the simulation components integrates the MatGL models with atomistic packages such as ASE and LAMMPS to perform molecular dynamics simulations.}
    \label{fig:matgl_workflow}
\end{figure}

MatGL is organized around four components: data pipeline, model architectures, model training and simulation interfaces. Fig.~\ref{fig:matgl_workflow} gives an overview of MatGL architecture, and detailed descriptions of each component are provided in the following subsections.

\section{Data Pipeline and Preprocessing}
The MatGL data pipeline consists primarily of \shellcmd{MGLDataset}, a subclass of \shellcmd{DGLDataset}, and \shellcmd{MGLDataLoader}, a wrapper around DGL's \shellcmd{GraphDataLoader}. \shellcmd{MGLDataset} is used for processing, loading and saving materials graph data, and includes tools to easily convert Pymatgen \shellcmd{Structure} or \shellcmd{Molecule} objects into directed or undirected graphs, while \shellcmd{MGLDataLoader} batches a set of preprocessed inputs with customized collate functions for training and evaluation. The main features of \shellcmd{MGLDataset} and \shellcmd{MGLDataLoader} are summarized in the following subsections.

\subsection{MGLDataset}
An important feature of \shellcmd{MGLDataset} is to provide a pipeline for processing graphs from inputs, loading and saving DGL graphs and labels. The commonly used inputs consist of the following items:
\begin{itemize}
    \item \shellcmd{structures}: A set of Pymatgen \shellcmd{Structure} or \shellcmd{Molecule} objects.
    \item \shellcmd{converter}: A graph converter that transforms a configuration into a DGL graph.
    \item \shellcmd{cutoff}: A cutoff radius that defines a bond between two atoms.
    \item \shellcmd{labels}: A list of target properties used for training.
\end{itemize}

Other inputs such as global state attributes and a cutoff radius for three-body interactions are optional depending on the model architecture and applications. The default units for PES properties are \si{\angstrom} for distance, eV for energy, eV \si{\angstrom}$^{-1}$ for force, and GPa for stress. 

\shellcmd{MGLDataset} also includes the ability to cache pre-processed graphs, which can facilitate the reuse of data for the training of different models. Once the \shellcmd{MGLDataset} is successfully loaded or constructed, the dataset can be randomly split into the training, validation, and testing sets using the DGL \shellcmd{split\_dataset} method. \shellcmd{MGLDataLoader} is then used to batch the separated training, validation and optional testing sets for either training or evaluation via PL modules.

\section{Model Architectures}

All GNN model architectures are implemented in the \shellcmd{matgl.models} package, using different layers implemented in the \shellcmd{matgl.layers} package. The models and layers are all subclasses of \shellcmd{torch.nn.Module}, which offers forward and backward functions for inference and calculation of the gradient of the outputs with respect to the inputs via the \shellcmd{autograd} function. Different models will utilize different combinations of layers, but, where possible, layers are implemented in a modular manner such that they are usable across different models (e.g., the MLP layer implementing a simple feed-forward neural network). 
MatGL offers various pooling operations, including set2set \cite{vinyals2015order}, average, and weighted average, to combine atomic, edge, and global state features into a structure-wise feature vector for predicting intensive properties. The pooled structural feature vector is then passed through an MLP for regression tasks, while a sigmoid function is applied to the output for classification tasks.

Table \ref{tab:architectures} summarizes the GNN models currently implemented in MatGL. The details of the models were already comprehensively described in the provided references, and interested readers are referred to those works. It should be noted that this is merely an initial set of model implementations.

\begin{table}[ht]
    \centering
    \begin{tabular}{ccp{6cm}ccc}
        \textbf{Name} & \textbf{Type} & \textbf{Brief Description} & \multicolumn{2}{c}{\textbf{Function}} & \textbf{Ref}\\
         &  &  & Prop. Pred. & MLIP & \\
        \hline
        MEGNet & Invariant & GNN with global state vector. & Yes & No & \citenum{chenGraphNetworksUniversal2019}\\
        M3GNet & Invariant & Extension of MEGNet with 3-body interactions. Used to implement the first uMLIP as well as property models. &  Yes & Yes &  \citenum{chenUniversalGraphDeep2022}\\
        CHGNet & Invariant & GNN with regularization of node features using magnetic moments from DFT. &  No & Yes &\citenum{dengCHGNetPretrainedUniversal2023}\\
        TensorNet & Equivariant & O(3)-equivariant GNN using Cartesian tensor representations, which is more computationally efficient compared to higher-rank spherical tensor models. &  Yes & Yes &\citenum{simeon2024tensornet}\\
        SO3Net & Equivariant & Minimalist SO(3)-equivariant GNN based on the spherical harmonics and Clebsch-Gordan tensor product. &  Yes & Yes &\citenum{schuttSchNetPack20Neural2023}\\
    \end{tabular}
    \caption{GNN architectures currently implemented in MatGL.}
    \label{tab:architectures}
\end{table}

In addition, all MatGL models subclass the \shellcmd{MatGLModel} abstract base class, which specifies that all models should implement a convenience \shellcmd{predict\_structure} method that takes in a Pymatgen Structure/Molecule and returns a prediction.

A key assumption in MLIPs is that the total energy can be expressed as the sum of atomic contributions. For PES models, the graph-convoluted atomic features are fed into either gated or equivariant gated multilayer perceptrons to predict the atomic energies. In addition, we have implemented a \shellcmd{Potential} class in the \shellcmd{matgl.apps.pes} package that acts as a wrapper to handle MLIP-related operations. For instance, a best practice for MLIPs is to first carry out a scaling of the total energies, for example, by computing either the formation energy or cohesive energy using the energies of the elemental ground state or isolated atom, respectively, as the zero reference. The \shellcmd{Potential} class takes care of accounting for the normalization factor in the total energies, as well as computing the gradient to obtain the forces, stresses and hessians. Other atomic properties such as magnetic moments and partial charges can also be predicted at the same time with the \shellcmd{Potential} class. 

\section{Training}
\begin{figure}[H]
    \centering
    \includegraphics[width=0.8\textwidth]{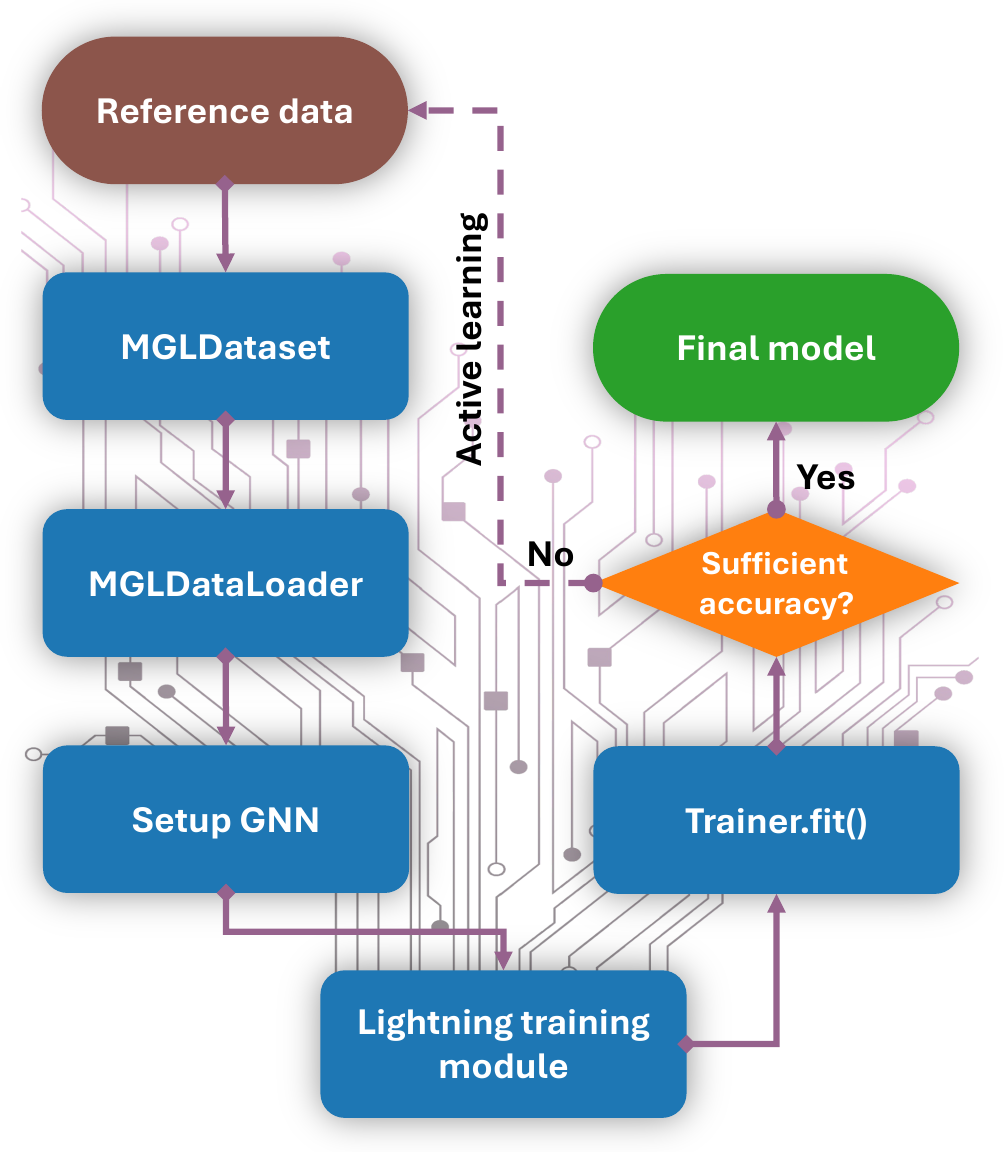}
    \caption{\textbf{Workflow for Training Property Models and Machine Learning Interatomic Potentials in MatGL.} The initial raw data includes a list of Pymatgen Structure/Molecule objects, optional global state attributes and labels such as structure-wise and PES properties. These inputs are used to preprocess training, validation and optional test sets containing a tuple of DGL graphs, labels, optional line graphs and state attributes using \shellcmd{MGLDataset}. These datasets are then fed into \shellcmd{MGLDataLoader} to create the batched inputs including graphs, state attributes and labels for training and validation. The GNN architecture is initialized with chosen hyperparameters and passed as inputs to PL training modules with training and validation data loaders. }
    \label{fig:training_workflow}
\end{figure}

The training framework for MatGL was built upon PL, which supports different efficient parallelization schemes and a variety of hardware including CPUs, GPUs and TPUs. MatGL provides two different PL modules including \shellcmd{ModelLightningModule} and \shellcmd{PotentialLightningModule} for property model and PES model training, respectively. Fig.~\ref{fig:training_workflow} illustrates the training workflow for building property models and MLIPs in MatGL. A set of reference calculations including structures and target properties is generated using ab initio methods and experiments. The reference structures are converted into a list of Pymatgen Structure/ Molecule objects, and target properties are stored in a dictionary, where the property names are the keys and corresponding values denote items. These inputs are passed through \shellcmd{MGLDataset}, followed by splitting the dataset into training, validation, and optional test sets, and then \shellcmd{MGLDataLoader} to obtain batched graphs, stacked state attributes, and labels. The desired GNN model architecture is initialized with requisite settings such as the number of radial basis functions, cutoff radii, etc. Various algorithms such as Glorot\cite{glorotUnderstandingDifficultyTraining2010} and Kaiming\cite{heDelvingDeepRectifiers2015} implemented in Pytorch can also be used to initialize the learnable parameters in GNNs.

The PL training modules include the \shellcmd{PotentialLightningModule} and \shellcmd{ModelLightningModule}. The major difference between the two modules is that the loss function for the  \shellcmd{PotentialLightningModule} is defined as a weighted sum of the errors of PES properties such as the energies, forces and stresses, and optionally, other atomic properties, such as magnetic moments and charges that affect the PES.

\subsection{Simulation Interfaces}
MatGL currently provides interfaces to the Atomistic Simulation Environment (ASE) and Large-scale Atomic/Molecular Massively Parallel Simulator (LAMMPS) to perform simulations with \shellcmd{Potential} models, i.e., MLIPs. For ASE, a \shellcmd{PESCalculator} class, initialized using a \shellcmd{Potential} class and state attributes, calculates energies, forces, stresses, and other atomic properties such as magnetic moments and charges for an ASE \shellcmd{Atoms} object, with the necessary conversion into DGL graphs being handled within the class itself. In addition, a \shellcmd{Relaxer} class allows users to perform structural optimization with different settings such as optimization algorithms (e.g. FIRE\cite{bitzekStructuralRelaxationMade2006}, BFGS\cite{broydenLocalSuperlinearConvergence1973, liu1989limited} and Gaussian process minimizer (GPMin)\cite{garijodelrioLocalBayesianOptimizer2019}) and variable cell relaxation for both Pymatgen \shellcmd{Structure/Molecule} and ASE \shellcmd{Atoms} objects. Finally, a \shellcmd{MolecularDynamics} class makes it easy to perform MD simulations under different ensembles with various thermostats such as Berendsen\cite{berendsenMolecularDynamicsCoupling1984}, Andersen\cite{andersenMolecularDynamicsSimulations1980}, Langevin\cite{schneiderMoleculardynamicsStudyThreedimensional1978} and Nos\'{e}-Hoover\cite{noseMolecularDynamicsMethod1984, hooverCanonicalDynamicsEquilibrium1985}. Additional functionality to compute material properties such as elasticity, phonon analysis and finding minimum energy paths using \shellcmd{PESCalculator} are available in the MatCalc\cite{Liu_MatCalc_2024} package. An interface to LAMMPS has also been implemented by AdvanceSoft, which utilizes \shellcmd{PESCalculator} to provide PES predictions for simulations. This interface enables the use of MatGL for a wide range of simulations supported by LAMMPS, including replica exchange\cite{sugitaReplicaexchangeMolecularDynamics1999} and grand canonical Monte Carlo (GCMC)\cite{adamsGrandCanonicalEnsemble1975}, etc. 

\subsection{Command-Line Interface}
MatGL offers a CLI for performing a variety of tasks including model training, evaluation and atomistic simulations. This interface minimizes the user's effort and time in preparing scripts to run calculations such as property prediction, geometry relaxation, MD, model training, and evaluation. 
\begin{itemize}
    \item \shellcmd{matgl predict}. This command is used to perform structure-wise property prediction, such as formation energy and band gap of materials. The prediction requires at least a structure file that can be read using the \shellcmd{Structure.from$\_$file} method from Pymatgen and a directory that stores the trained property model. Additionally, predictions for multiple structure-wise properties are also supported.
    \item \shellcmd{matgl relax}. This command is used to perform geometry relaxation using the \shellcmd{Relaxer} class with a trained MLIP. Users can flexibly decide whether to perform variable-cell relaxation and can adjust the maximum allowable force components to define the relaxation criteria. The default optimizer is the FIRE algorithm \cite{bitzekStructuralRelaxationMade2006}, although other optimization algorithms are also available.
    \item \shellcmd{matgl md}. This command is used to perform MD simulations using the \shellcmd{MolecularDynamics} class. Similar to \shellcmd{matgl relax}, it also requires a structure and a trained MLIP. Users can customize various simulation parameters, including the step size, ensemble type, number of time steps, target pressure, and temperature. Furthermore, ensemble-dependent settings such as collision probability, external stress, and coupling constants for thermostats can be also adjusted to specific systems. 
    \item \shellcmd{matgl train} and \shellcmd{matgl evaluate}. These commands are used to perform model training and evaluation, including data preprocessing, splitting, setting up the GNN architecture, and configuring Lightning modules. Users only need to provide an input file containing structures and their corresponding target properties, along with the settings for graph construction, GNN architecture, and training hyperparameters. These settings can be modified in the configuration file or specified as input arguments.
\end{itemize}

\section{Benchmarks}
In the following sections, we benchmark the performance of different GNN architectures, trained on various popular datasets, in terms of accuracy and inference time.

\subsection{Property Prediction}
This section summarizes the performance of various GNN architectures for predicting various properties of the QM9 molecular\cite{dunnBenchmarkingMaterialsProperty2020} and matbench bulk crystal\cite{ramakrishnanQuantumChemistryStructures2014a} datasets. 

\subsubsection{QM9}
The QM9 dataset contains 130,831 organic molecules including H, C, N, O and F. GNN models were trained on the isotropic polarizability ($\alpha$), free energy ($G$) and the gap ($\Delta \epsilon$) between the highest occupied molecular orbital (HOMO) and the lowest unoccupied molecular orbital (LUMO), which were computed with DFT with the B3LYP functional. 

Table~\ref{tab:qm9_mae} shows the MAE of different GNN architectures. Consistent with previous analyses, MEGNet obtains the highest errors, while other models are comparable. For example, MEGNet achieves validation and test MAEs of 0.037 eV for free energy, while other models reach a range of 0.025-0.027 eV. It should be noted that these experiments aim to demonstrate the capabilities of MatGL with consistent settings. For comparing the best accuracy between different architectures, an extensive search for preprocessing treatments of target properties and hyperparameters, such as learning rate, scheduler, and weight initialization, is required.

\begin{table}[ht]
    \centering
    \begin{tabular}{c|c|c|c}
    Model    & $\alpha$ ($a_{0}^{3}$ ) & $G$ (eV) & $\Delta \epsilon$ (eV) \\
    \hline
    MEGNet    &   0.066/0.113/0.114 &  0.032/0.037/0.037  & 0.031/0.079/0.081 \\
    M3GNet    &   0.040/0.089/0.087 &  0.019/0.025/0.025  & 0.014/0.059/0.061 \\
    TensorNet &   0.050/0.083/0.083 &  0.024/0.027/0.027  & 0.021/0.064/0.065 \\
    SO3Net    &   0.046/0.068/0.069 &  0.022/0.025/0.027  & 0.024/0.059/0.060 
    \end{tabular}
    \caption{\textbf{Mean absolute errors (MAEs) of GNN models trained on QM9 dataset.} Calculated MAEs of isotropic polarizability $\alpha$, free energy $G$ and HOMO–LUMO gap $\Delta \epsilon$ with MEGNet, M3GNet, TensorNet and SO3Net. The numbers are reported in the order of training/validation/test MAEs. The dataset was divided into training, validation, and test sets with a split ratio of 0.9, 0.05, and 0.05, respectively.} 
    \label{tab:qm9_mae}
\end{table}

\subsubsection{Matbench}

We trained four different GNNs on three properties: formation energy (\(E_{\mathrm{form}}\)), Voigt-Reuss-Hill bulk modulus (log($K_{\mathrm{vrh}}$)), and shear modulus (log($G_{\mathrm{vrh}}$)). The datasets contained 132,752, 10,987, and 10,987 crystals, respectively, resulting in a total of 12 property models.
\begin{table}[ht]
    \centering
    \begin{tabular}{c|c|c|c|c}
    Model    & $E_{\mathrm{form}}$ (eV/atom) & log($K_{\mathrm{VRH}}$) (log(GPa)) & log($G_{\mathrm{VRH}}$) (log(GPa)) & $E_{\mathrm{G}}$ (eV)\\
    \hline
    MEGNet    &   0.015/0.037/0.037 & 0.033/0.063/0.075  & 0.046/0.085/0.090 & 0.072/0.213/0.220\\
    M3GNet    &   0.007/0.020/0.020 & 0.039/0.054/0.065  & 0.032/0.081/0.091 & 0.032/0.160/0.170\\
    TensorNet &   0.008/0.024/0.024 & 0.031/0.054/0.060  & 0.046/0.082/0.090 & 0.043/0.163/0.177\\
    SO3Net    &   0.008/0.022/0.022 & 0.035/0.052/0.060 & 0.031/0.079/0.083 & 0.033/0.169/0.180
    \end{tabular}
    \caption{\textbf{Mean absolute errors (MAEs) of GNNs trained on Matbench dataset.} Calculated MAEs of formation energy $E_{\mathrm{form}}$, Voigt-Reuss-Hill bulk $K_{\mathrm{VRH}}$ and shear modulus $G_{\mathrm{VRH}}$ as well as bandgap with MEGNet, M3GNet, TensorNet and SO3Net. The numbers are reported in the order of training/validation/test MAEs. The dataset was divided into training, validation, and test sets with a split ratio of 0.9, 0.05, and 0.05, respectively.}
    \label{tab:matbench_mae}
\end{table}

Table~\ref{tab:matbench_mae} reports the MAEs of material properties including formation energy, bulk/shear modulus and bandgap with respect to reference DFT-PBE results. All GNN models achieve state-of-the-art accuracy in terms of training, validation and test errors\cite{dunnBenchmarkingMaterialsProperty2020, wangCompositionallyRestrictedAttentionbased2021a}. MEGNet generally obtains the highest MAEs compared to other models. For instance, the calculated validation and test MAEs of MEGNet for formation energy are 0.037 eV atom$^{-1}$, while other models significantly reduce the error by 40\%. The poor performance of MEGNet is attributed to the less informative geometric representation of structures based only on bond distances. Recent studies \cite{pozdnyakovIncompletenessGraphNeural2022} find that distance-only GNNs fail to uniquely distinguish atomic environments, which affects the accuracy of structure-wise properties due to degeneracies caused by the incompleteness of representation. Other models like M3GNet, TensorNet and SO3Net achieve considerably higher accuracy by taking additional geometric information, such as bond angles and relative position vectors, into account. The learning curves for QM9 and Matbench are provided in Fig. S1-S2.

\subsubsection{Inference time of property models}
\begin{table}[ht]
    \centering
    \begin{tabular}{c|c|c}
    Model    & QM9  & Matbench \\
    \hline
    MEGNet    &   11.996 &  11.137   \\
    M3GNet    &   19.715 &  20.089   \\
    TensorNet &   14.945 &  13.694   \\
    SO3Net    &   14.371 &  32.601   
    \end{tabular}
    \caption{\textbf{Inference times of GNN models for property prediction.} The numbers represent the inference times (in seconds) for MEGNet, M3GNet, TensorNet, and SO3Net on the QM9 (free energy) and Matbench (formation energy) test sets, which contain 6,541 and 6,637 structures, respectively. All property predictions were performed using a single Nvidia RTX 3090 and A6000 GPU for QM9 and Matbench, respectively.}
    \label{tab:prop_time}
\end{table}

Table~\ref{tab:prop_time} shows the inference time of the test set for the QM9 and Matbench datasets for the different GNN models. MEGNet achieves the shortest inference time with 12 s and 11 s for around 6,500 small molecules and crystals although the accuracy is the worst. TensorNet generally achieves the best compromise between accuracy and efficiency, taking less than 15 seconds for both datasets. M3GNet and SO3Net has the longest inference time for molecules and crystals, respectively. This shows that the SO3Net is slower than M3GNet when the number of neighbors within a spatial cutoff sphere is larger.
\subsection{Potential Energy Surface}
This section summarizes the performance of various GNN model architectures in constructing MLIPs using popular large databases such as the ANI-1x\cite{smithLessMoreSampling2018} and MPF-2021.2.8. The results and benchmarks are presented below.

\subsubsection{ANI-1x}

The first benchmark dataset is ANI-1x\cite{smithLessMoreSampling2018}, which contains roughly 5 million conformers generated from ~57,000 distinct molecules containing H, C, N, and O for constructing general-purpose organic molecular MLIPs. 
We also included the Transfer-Learning M3GNet (M3GNet-TL) MLIPs from the pre-training ANI-1xnr dataset\cite{zhangExploringFrontiersCondensedphase2024a} by adapting the pretrained embedded layer and only optimizing other model parameters for comparison. We noted that the ANI-1xnr dataset encompasses a significantly larger configuration space compared to ANI-1x, owing to the extensive structural diversity obtained from condensed-phase reactions. These reactions include carbon solid-phase nucleation, graphene ring formation from acetylene, biofuel additive reactions, methane combustion, and the spontaneous formation of glycine from early earth small molecules.

Table~\ref{tab:ani_1x_mae} shows the MAEs of energies and forces computed with different GNNs with respect to DFT. Both M3GNet and TensorNet achieve comparable training and validation MAEs of energies and forces, while SO3Net significantly outperforms them. A similar conclusion can be drawn from the test errors showing that SO3Net achieves the lowest MAE in terms of energies and forces.

The results are consistent with previous findings, indicating that equivariant models are typically more accurate and transferable than invariant models for molecular systems. Moreover, M3GNet-TL reduces the errors in energies and forces by 10 to 15\% compared to M3GNet trained from scratch and also exhibits significantly faster convergence, as shown in Fig. S3. The improvements are attributed to the pre-trained embedded layer from ANI-1xnr dataset that covers a greater diversity of local atomic environments.

\begin{table}[ht]
    \centering
    \begin{tabular}{c|c|c}
    Model & Energy (meV atom$^{-1}$) & Force (eV $\si{\angstrom}^{-1}$) \\
    \hline
    M3GNet     & 4.565/4.592/3.746  & 0.092/0.093/0.085 \\
    M3GNet-TL & 3.923/3.968/3.381 & 0.081/0.082/0.075 \\
    TensorNet  & 4.424/4.448/3.015  & 0.088/0.088/0.074 \\
    SO3Net     & 2.281/2.286/1.596 &  0.046/0.046/0.035 
    \end{tabular}
    \caption{\textbf{Mean absolute errors on ANI-1x subset.} The numbers are the calculated energy and force errors of M3GNet, TensorNet, and SO3Net compared to DFT. The "M3GNet-TL" indicates the transfer learning from the pre-trained M3GNet model on ANI-1xnr dataset. The numbers are listed in the order of training, validation, and test. The dataset was divided into training, validation, and test sets with a split ratio of 0.9, 0.05, and 0.05, respectively.} 
    \label{tab:ani_1x_mae}
\end{table}

\subsubsection{Extrapolation to COMP6 dataset}
\begin{figure}[H]
    \centering
    \includegraphics[width=1.0\textwidth]{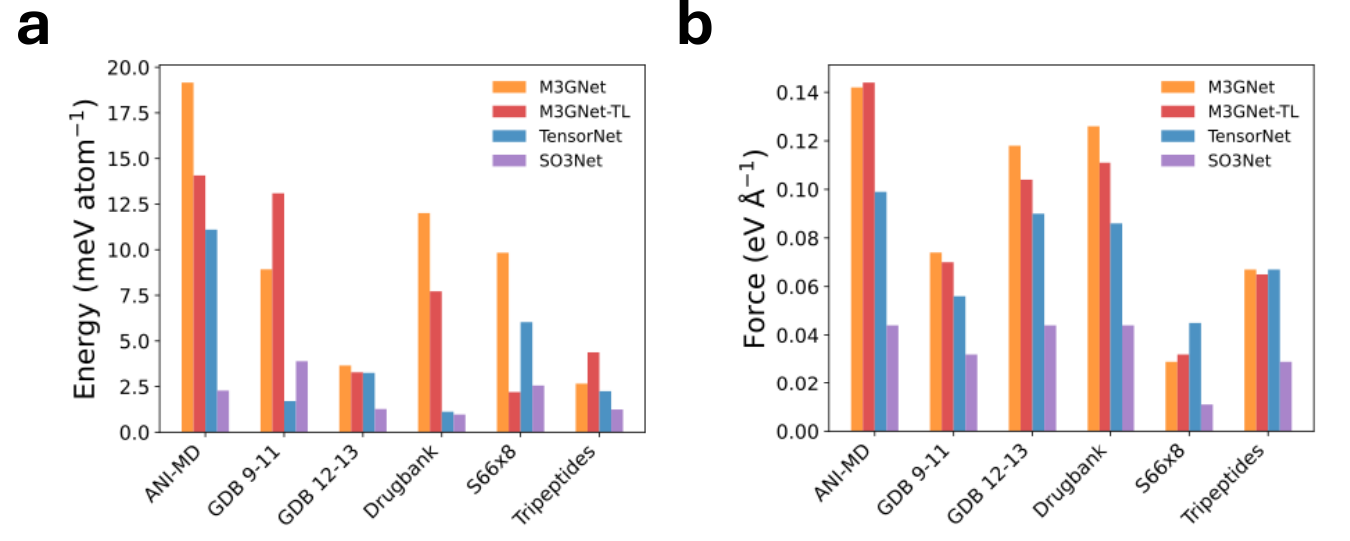}
    \caption{\textbf{Mean absolute errors on COMP6 benchmark.} The bar plot of \textbf{a} energy and \textbf{b} force errors for M3GNet, transfer-learning M3GNet (M3GNet-TL) from ANI-1xnr, TensorNet and SO3Net with respect to DFT. }
    \label{fig:COMP6}
\end{figure}

To further evaluate the extrapolation abilities of GNN models, we compare the energies and forces on the molecules obtained from COMP6 benchmarks with respect to DFT. Table~\ref{fig:COMP6} shows the MAE of energies and forces computed with M3GNet, M3GNet-TL, TensorNet and SO3Net. Both M3GNet and M3GNet-TL perform the worst in terms of energy and force errors above 14 meV atom$^{-1}$ and 0.14 eV $\si{\angstrom}^{-1}$ on the ANI-MD dataset, which comprises molecular dynamics (MD) trajectories of 14 well-known drug molecules and 2 small proteins. The large errors may be attributed to the poor transferability of MLIPs trained on small molecules to larger ones, as the largest molecule in the training set contains 63 atoms, whereas the molecules in the ANI-MD dataset have 312 atoms. The TensorNet significantly reduces the error of energies and forces to 11 meV atom$^{-1}$ and 0.1 eV $\si{\angstrom}^{-1}$, while SO3Net further reduces to 2.3 meV atom$^{-1}$ and 0.044 eV $\si{\angstrom}^{-1}$. This trend can be also found in other benchmark datasets. 

\begin{figure}[H]
    \centering
    \includegraphics[width=1.0\textwidth]{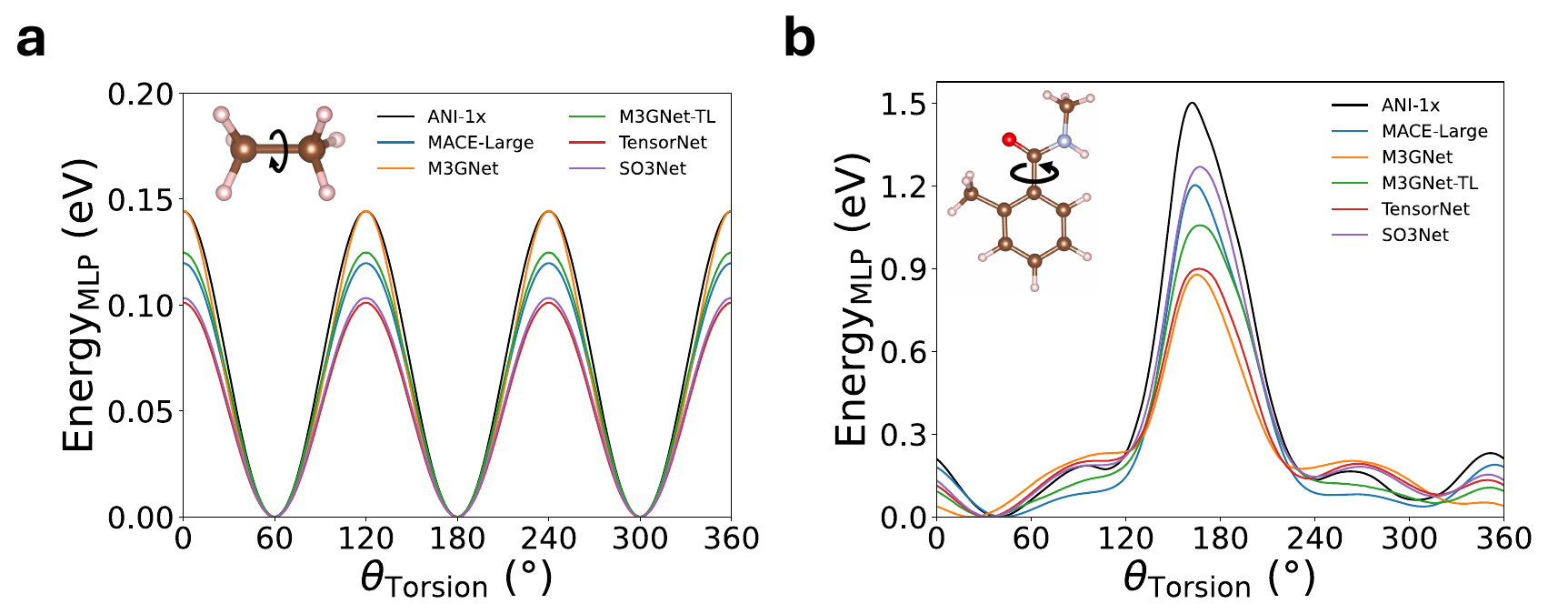}
    \caption{\textbf{Potential energy surface of organic molecules during torsion.} The torsion energy profile of \textbf{a} ethane and \textbf{b} dimethyl-benzamide were computed with different MLIPs. The reference ANI-1x\citenum{smithLessMoreSampling2018} and MACE-Large\citenum{kovacs2023mace} were plotted in black and purple lines. The black arrows indicate the dihedral torsion of molecules. }
    \label{fig:mole_tors}
\end{figure}

To further demonstrate the performance of constructed MLIPs from MatGL with state-of-the-art models, we calculated the energy of two well-known molecules with respect to the dihedral torsion. Fig.~\ref{fig:mole_tors}a shows the PES of ethane during torsion. All MLIPs, including reference ANI-1x\cite{smithLessMoreSampling2018} and MACE-Large\cite{kovacs2023mace}, predict the same torsion angles for the maxima and minima of the PESs, while the energy barriers are slightly different. For instance, both ANI-1x and M3GNet predict a higher energy barrier of 0.15 eV, whereas MACE-Large and M3GNet-TL obtain 0.125 eV. SO3Net and TensorNet predict the lowest energy barrier of 0.1 eV. For the case of a more complex di-methyl-benzamide molecule, all the MLIPs provide a similar shape of PESs with respect to different dihedral angles. Still, the predicted barrier heights are different. For example, the ANI-1x model has the largest barrier height of 1.5 eV at 180\textdegree, while both TensorNet and M3GNet considerably underestimate the energy barrier by 0.6 eV. The energy barriers for M3GNet-TL, SO3Net, and MACE-Large range from 0.9 to 1.2 eV.

\subsubsection{Materials Project  MPF.2021.2.8 database}
The second dataset is the manually selected subset of MPF.2021.2.8. All dataset, which includes all geometry relaxation trajectories from both the first and second step calculations in the Materials Project.  The total number of crystal structures is 185,877. Moreover, the isolated atoms of 89 elements were also included in the training set to improve the extrapolability of the final potential. The details of data generation and selection can be found in ref.\cite{qi2024robust}.
Here we excluded SO3Net from the benchmarks due to its relatively high sensitivity to noisy datasets, which led to extremely large fluctuations in training errors. 

Table~\ref{tab:mp_mae} shows that CHGNet generally outperforms M3GNet and is noticeably better than TensorNet in terms of energies, forces and stresses. The convergence of validation loss and PES properties was plotted in Fig. S4. This can be attributed to the fact that the CHGNet provides additional message passing between angles and edges compared to M3GNet. Moreover, the DFT calculation settings, such as electronic convergence and grid density in reciprocal space, are less strict, resulting in large numerical noise in forces and stresses, which makes the training particularly challenging for equivariant models that are very sensitive to these properties. Furthermore, most structures are crystals without complicated structural diversity, which reduces the strength of equivariant models in providing a more informative representation of complex atomic environments. More detailed benchmarks on structurally diverse datasets with stricter electronic convergence for constructing general-purpose universal MLIPs are required in future studies.
\begin{table}[ht]
    \centering
    \begin{tabular}{c|c|c|c}
     Model    & Energy (meV atom$^{-1}$) & Force (eV $\si{\angstrom}^{-1}$) & Stress (GPa)  \\
     \hline
      M3GNet   &  19.817/22.558/23.037 & 0.063/0.072/0.071 & 0.259/0.399/0.351 \\
      TensorNet & 28.628/29.708/30.313 & 0.078/0.083/0.083 & 0.361/0.471/0.394 \\
      CHGNet & 17.256/18.226/19.897 & 0.054/0.061/0.061 & 0.254/0.347/0319
    \end{tabular}
    \caption{\textbf{Mean absolute error on MPF-2021.2.8 subset.} The numbers are the calculated energy, force and stress mean absolute errors (MAEs) of M3GNet, TensorNet, and CHGNet compared to DFT. The numbers are listed in the order of training, validation, and test. The dataset was divided into training, validation, and test sets with a split ratio of 0.9, 0.05, and 0.05, respectively.} 
    \label{tab:mp_mae}
\end{table}
We also performed benchmarks on crystals, particularly focusing on binary systems obtained from the Materials Project database. 
\begin{figure}[H]
    \centering
    \includegraphics[width=1.0\textwidth]{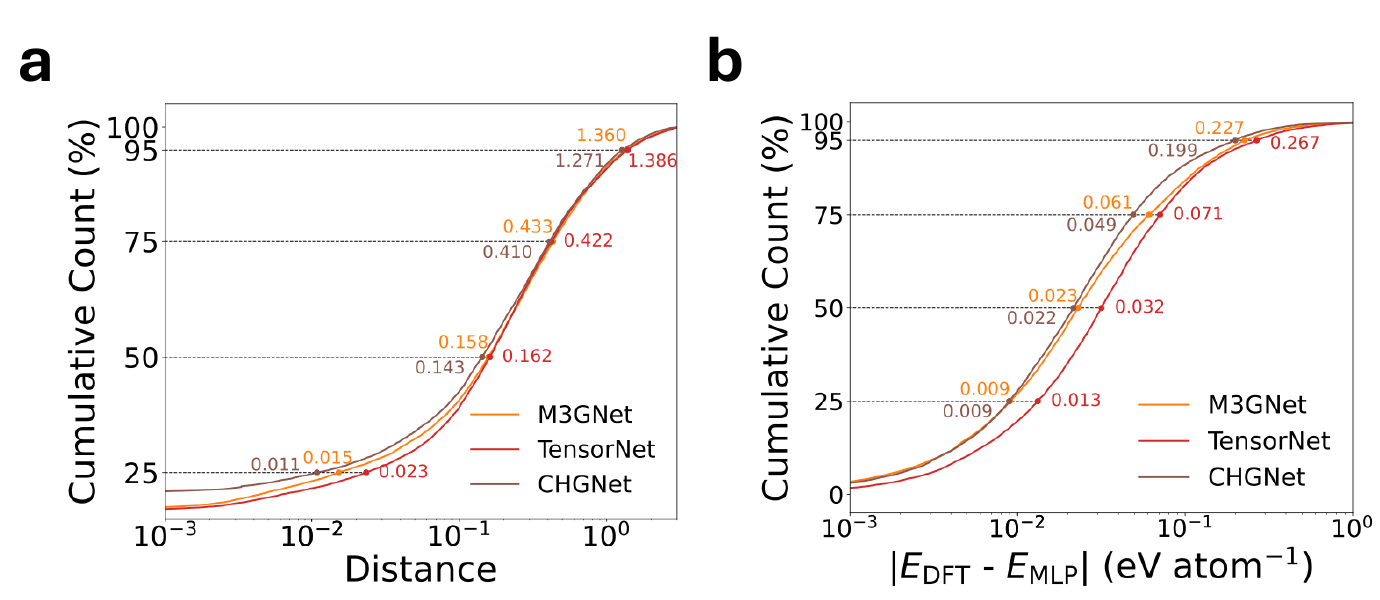}
    \caption{\textbf{Performance of universal potentials for variable-cell geometry relaxation of binary crystals.} \textbf{a} Cumulative absolute fingerprint distance of DFT and MLIP relaxed structures using CrystalNN algorithm, and \textbf{b} Cumulative absolute errors of DFT and MLIP energies of relaxed crystals. }
    \label{fig:rmsd_and_energy}
\end{figure}
The first step is to investigate the performance of GNNs on the geometry relaxation of binary crystals and corresponding energies with respect to DFT. It should be noted that such benchmarks for existing uMLIPs have been reported in recent studies\cite{gonzales2024benchmarking,yu2024systematic}.Fig.~\ref{fig:rmsd_and_energy}a shows the cumulative structural fingerprint distance between DFT and MLIP relaxed structures using CrystalNN algorithm\cite{panBenchmarkingCoordinationNumber2021}, which indicates the similarity between the two structures based on the local atomic environments. Overall, both M3GNet and TensorNet have similar performance in terms of fingerprint distance. CHGNet only shows a modest improvement, with more structures within a distance of about 0.01 compared to M3GNet and TensorNet. Fig.~\ref{fig:rmsd_and_energy}b shows the cumulative absolute energy errors of MLIPs with respect to DFT. CHGNet predicts that about 60\% of structures have an energy difference below 25 meV atom$^{-1}$. This is comparable to M3GNet and 10\% better than TensorNet.

\begin{figure}[H]
    \centering
    \includegraphics[width=1.0\textwidth]{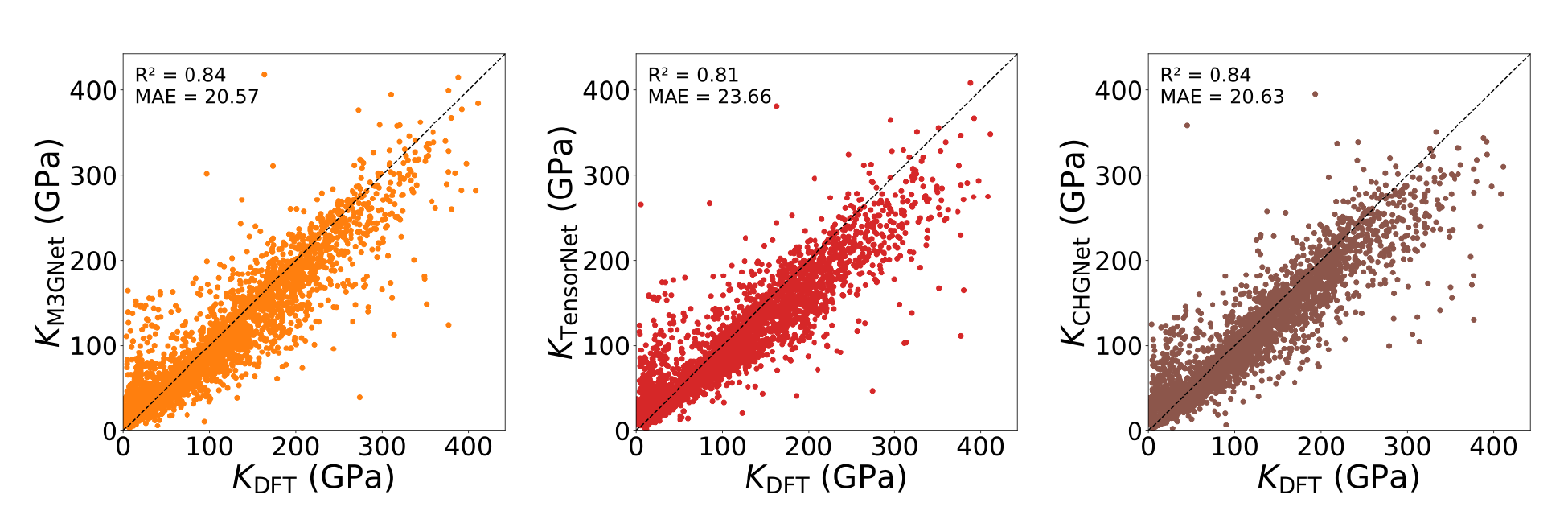}
    \caption{\textbf{Performance of universal potentials for bulk modulus of binary crystals.} Parity plots for Voigt-Reuss-Hill bulk modulus calculated with M3GNet, TensorNet and CHGNet compared to DFT. }
    \label{fig:bulk_modulus}
\end{figure}
We also compared the predicted bulk modulus with different models. Fig.~\ref{fig:bulk_modulus} shows the parity plots of bulk modulus computed with universal MLIPs and DFT. All models have similar R$^{2}$ scores and MAEs, reaching 0.8 and 20 GPa. 

\begin{figure}[H]
    \centering
    \includegraphics[width=1.0\textwidth]{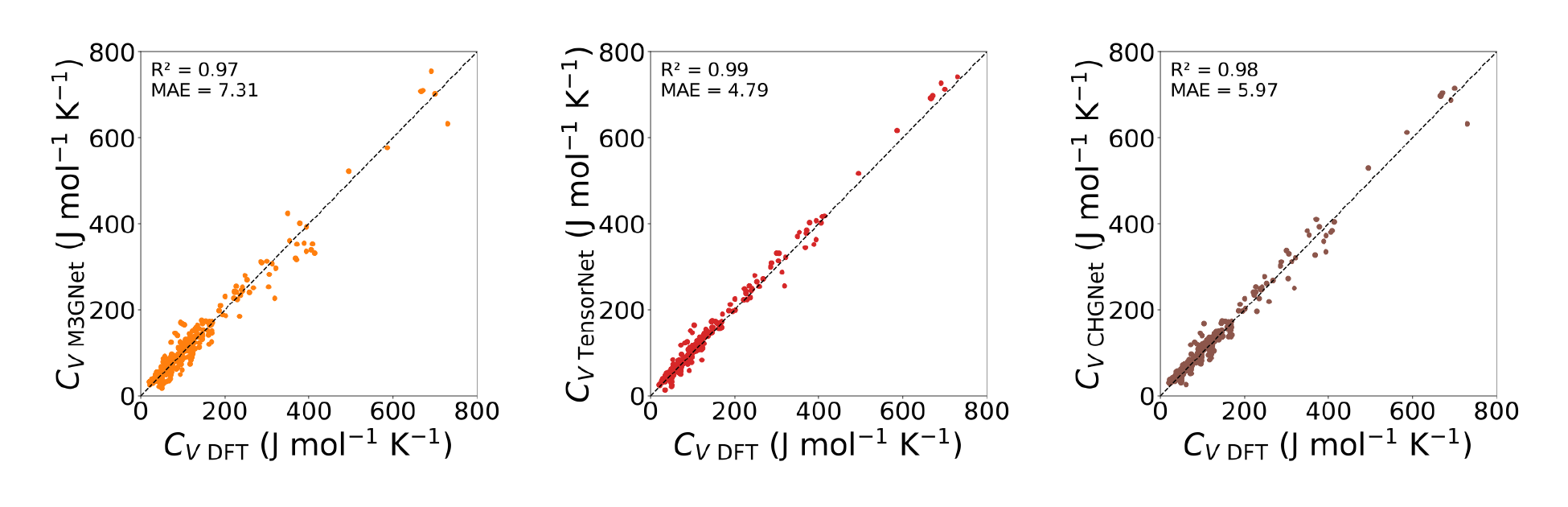}
    \caption{\textbf{Comparison of universal potentials for the heat capacity of binary crystals.} Parity plots for heat capacity calculated with M3GNet, TensorNet and CHGNet compared to DFT.} 
    \label{fig:heat_capacity}
\end{figure}

Finally, we computed the heat capacity of binary systems at 300K under phonon harmonic approximation and compared the results with DFT reference data at the PBEsol level obtained from Phonondb. Fig.~\ref{fig:heat_capacity} shows that all models are in very good agreement with DFT. A very recent study \cite{loew2024universal} noted a small shift between PBE and PBE-sol on the prediction of phonon properties.
Nevertheless, these benchmarks demonstrate that our trained MLIPs can provide a preliminary reliable prediction on material properties by performing geometry relaxations and phonons. These uMLIPs can perform reasonably stable MD simulations across a wide range of systems at low temperatures, as their covered configuration space partially overlaps with relaxation trajectories near the equilibrium region\cite{batatia2023foundation,chenUniversalGraphDeep2022,dengCHGNetPretrainedUniversal2023}.

\subsubsection{Inference time of MLIPs}
\begin{figure}[H]
    \centering
    \includegraphics[width=1.0\textwidth]{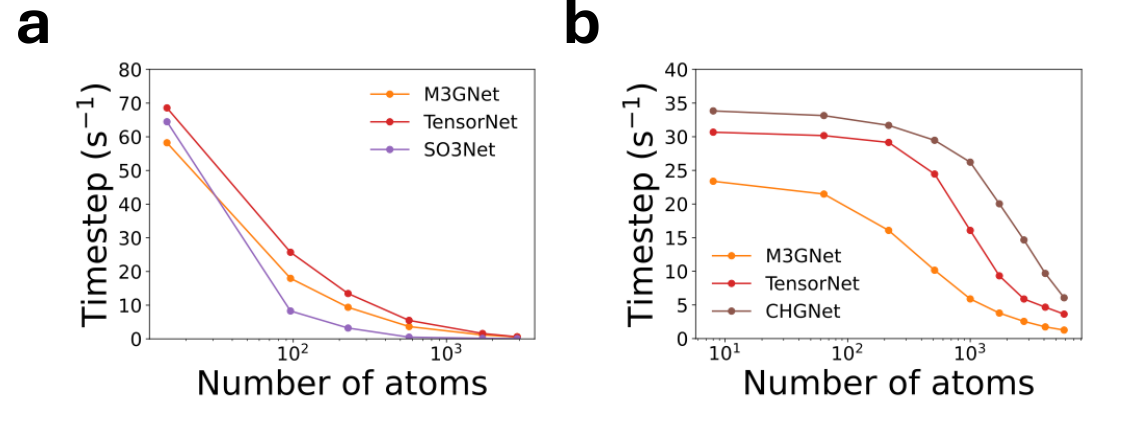}
    \caption{\textbf{Inference time of MD simulations.} The number of timesteps per second for \textbf{a} \textit{NVT} simulations of water clusters with different sizes using ASE and \textbf{b} \textit{NPT} simulations of various silicon-diamond supercells using LAMMPS is reported. All MD simulations were performed using a single Nvidia RTX A6000 GPU.}
    \label{fig:inference_md}
\end{figure}
The reliability of material properties extracted from MD simulations critically depends on the accuracy of trained MLIPs. MatGL provides ASE and LAMMPS interfaces to perform MD simulations, enabling the benchmarking of different GNN architectures\cite{fu2023forces,bihani2024egraffbench}.
In addition to the accuracy of GNNs, computational efficiency is crucial for large-scale atomistic simulations. We used the above MLIPs to perform MD simulations with 1000 timesteps for scalability tests with a single GPU via ASE and LAMMPS interfaces. Fig.~\ref{fig:inference_md}a shows the computational time for \textit{NVT} simulations of non-periodic water clusters using ASE, with increasing sizes from 15 to 2892 atoms. SO3Net becomes significantly more demanding than TensorNet and M3GNet when simulating clusters with more than 100 atoms. TensorNet is the most efficient for all cases compared to M3GNet and SO3Net due to its model architecture, which does not require costly three-body calculations and tensor products. With a more scalable and optimized LAMMPS interface, Fig.~\ref{fig:inference_md}b shows the computational time of \textit{NPT} simulations for silicon diamond supercells ranging from 8 to 5832 atoms, where each Si atom contains around 70 neighbors within a spatial cutoff of 5 $\mathrm{\si{\angstrom}}$. CHGNet achieves the shortest computational time, while the computational cost of M3GNet is the highest. This is likely due to the additional cost of a larger cutoff for counting triplets and three-body interactions. These models can already serve as a "foundation" model for preliminary calculations with reasonably good accuracy. Moreover, building customized MLIPs often requires extensive AIMD simulations to sample the snapshots from the trajectories for training. Such demanding AIMD simulations can be replaced by the universal MLIPs with considerably reduced costs\cite{qi2024robust}.

\section{Discussion}
Graph deep learning has made tremendous progress in atomistic simulations. Here we have implemented MatGL, which covers four major components including data-pipelines, state-of-the-art graph deep learning architectures, Pytorch-Lightning training modules, interfaces with atomistic simulation packages, and command-line interfaces. 
We also provided detailed documentation and examples to help users become familiar with training their custom models and conducting simulations using ASE and LAMMPS packages in our public Github repository. In addition, we provided multiple pretrained models, including 28 for structural properties and 6 for foundational MLIPs, applicable to organic molecules and materials with reliable accuracy. With the combination of excellent chemical scalability and large databases, these models empower users to perform simulations across a wide range of applications, speeding up materials discovery by enabling high-throughput screening of hypothetical materials across a large chemical space\cite{chen2024accelerating,ojih2024graph,sivak2024discovering,taniguchi2024exploration}.
Furthermore, users can efficiently train their customized models with significantly faster convergence through fine-tuning from our available pretrained models. Additionally, MatGL allows developers to design their own graph deep learning architectures and benchmark their performance with minimum effort, complimented by the modules available in the library.
MatGL has been integrated into various frameworks, including MatSciML\cite{miret2023the} and the Amsterdam Modeling Suite\cite{ADF2001}, expanding access for researchers in materials science and chemistry to conduct computational studies on a wide range of materials using GNNs.
In future work, the efficiency of MLIPs can be further enhanced by integrating multi-GPU support with efficient parallelization algorithms\cite{parkScalableParallelAlgorithm2024a}. Besides, training on massive databases exceeding millions of structures may encounter bottlenecks due to the memory needed to store all graphs and labels. To address this, the lightning memory-mapped database can be utilized to manage such large-scale training with affordable computational resources. We expect that the upcoming version of MatGL will substantially increase the accessible training set size for constructing foundation models and enhance the efficiency of large-scale MD simulations, enabling the study of many interesting phenomena in materials science and chemistry.

\section{Methods}

\subsection{Model Training}
All models were trained using \shellcmd{PotentialLightningModule} for structure-wise properties and \shellcmd{ModelLightningModule} for potential energy surfaces (PESs). The optimizer was chosen to be the AMSGrad variant of AdamW with a learning rate of 10$^{-3}$. The weight decay coefficient was set to 10$^{-5}$. The cosine annealing scheduler was used to adjust the learning rate during the training. The maximum number of iterations and minimum learning rate were set to 10$^{4}$ and 10$^{-5}$, respectively. The mean absolute error of predicted and target properties was selected to calculate the loss function. The additional relative importance of energies, forces and stresses (1:1:0.1) was introduced for PES training. The maximum number of epochs was set to 1000, and early stopping was achieved with the patience of 500 epochs. The gradient for model weight updates was accumulated over 4 batches, and the gradient clipping threshold to prevent gradient explosion was set to 2.0. A full table of hyperparameters for each model and training module is provided in Table S3-S7. 
For detailed descriptions of all models, the interested readers are referred to the respective publications.

\subsection{Benchmarking}

\subsubsection{Dihedral Torsion}
The initial structures of ethane and dimethylbenzamide were relaxed using the FIRE algorithm with molecular MLIPs under a stricter force threshold of 0.01 eV $\si{\angstrom}^{-1}$. The conformers for scanning the dihedral angles were generated using RDKit\cite{landrum2013rdkit} at 1$^{\circ}$  intervals, resulting in a total of 359 single-point calculations to produce the PES.

\subsection{Geometry Relaxation of Binary crystals}
The 20160 initial DFT-relaxed binary crystals were taken from the Materials Project database. All these structures were re-optimized using universal MLIPs with variable cell geometry relaxation within a lighter force threshold of 0.05 eV $\si{\angstrom}$. The default settings for CrystalNN were employed to measure the similarity between the DFT and MLIP-relaxed structures based on the fingerprints of their local environments. It should be noted that two structures failed during relaxation with CHGNet due to the failed construction of bond graphs caused by unphysical configurations.

\subsection{Voigt-Reuss-Hill Bulk Modulus and Heat Capacity}
A total of 4,653 and 1,183 binary crystals with available Voigt-Reuss-Hill bulk modulus and heat capacity data were obtained from the Materials Project and PhononDB, respectively. Additional filters were applied to unconverged DFT calculations and unphysical bulk modulus and the remaining 3576 structures finally were analyzed. As for heat capacity, 1183 binary crystals were compared.
All predicted properties derived from MLIPs were calculated using ElasticityCalc and PhononCalc from the MatCalc library. The default settings were used, except for a stricter force convergence threshold of 0.05 eV/Å. Notably, all phonon calculations were completed successfully with the lighter symmetry search tolerance set to 0.1.
 
\subsection{Dataset details}
All datasets except ANI-1x were randomly split into training, validation and test sets with a ratio of 0.9, 0.05 and 0.05, respectively. Due to the large size of the ANI-1x dataset, only a subset was used for demonstration purposes. We randomly sample the conformations of each molecule with the ratio of 0.2, 0.05, and 0.05 for training, validation and testing. With the molecules containing less than 10 conformations, all conformations are included in the training to ensure that every molecule in the ANI-1x dataset is included in the training set. The description of datasets was summarised in the following subsection.

\subsubsection{QM9}
QM9 consists of 130,831 organic molecules including H, C, M, O, F. It is a subset of GDB-17 database\cite{ruddigkeitEnumeration166Billion2012a} for isotropic polarizability, free energy and the gap between HOMO and LUMO were calculated using DFT at the level of B3LYP/6-31G. 

\subsubsection{Matbench}
The Matbench dataset consists of 132,752 and 10,987 crystals for formation energy and bulk/shear modulus computed with DFT, respectively. All datasets were generated using the Materials Project API on 4/12/2019. The details can be found in ref.\cite{dunnBenchmarkingMaterialsProperty2020}

\subsubsection{ANI-1x}
The ANI-1x is the extension of ANI-1 dataset\cite{smithLessMoreSampling2018} by performing active learning based on three different samplings including molecular dynamics, normal mode and torsion. All energies and forces of conformers are calculated using DFT at wB97x/6-31G level.

\subsection{M3GNet-MS}
The M3GNet-MS dataset consists of 185,877 configurations sampled manually in the relaxation trajectories of 60,000 crystals from Materials Project. Additionally, 89 different isolated elements were also included in the training set

\section{Data Availability}
All datasets used in this work are publicly available in the following links:\newline
QM9: \href{QM9}{https://doi.org/10.6084/m9.figshare.c.978904.v5}\newline
Matbench: \href{Matbench}{https://hackingmaterials.lbl.gov/automatminer/datasets.html}\newline
ANI-1x: \href{ANI-1x}{https://doi.org/10.6084/m9.figshare.c.4712477.v1}\newline
ANI-1xnr: \href{ANI-1xnr}{https://doi.org/10.6084/m9.figshare.22814579}\newline
COMP6: \href{COMP6}{https://github.com/isayev/COMP6}\newline
MPF-2021.2.8: \href{MPF-2021.2.8}{https://figshare.com/articles/dataset/20230723\_figshare\_DIRECT\_zip/23734134}

\section{Code Availability}
All implementations are available in MatGL(\href{MatGL}{https://github.com/materialsvirtuallab/matgl}). The pretrained models will be provided in the latest released version of MatGL.

\section{Contribution}
S.P.O. and S.M. conceived the idea and initiated the research project. T.W.K. led the implementation of major components with support and advice from S.P.O.. T. W. K. also contributed to most of the model training and benchmarking. B.D. contributed to the implementation and training of CHGNet and improved some parts of implementations in MatGL. M.N. contributed to the preliminary implementation of MEGNet. L. B. helped with the implementation of the CHGNet and graph construction. J.Q. helped with the implementation of graph construction and the training of MLIPs. R.L. contributed to the design of workflow and benchmarking for different GNN models trained by T.W.K., J. Q. and B.D.. E.L. helped with the implementation of different basis functions. T. W. Ko and S. P. Ong wrote the initial manuscript and all authors contributed to the discussion and revision.

\begin{acknowledgement}
This work was intellectually led by the U.S. Department of Energy, Office of Science, Office of Basic Energy Sciences, Materials Sciences and Engineering Division under contract No. DE-AC02-05-CH11231 (Materials Project program KC23MP). This research used resources of the National Energy Research Scientific Computing Center (NERSC), a Department of Energy Office of Science User Facility using NERSC award DOE-ERCAP0026371. T. W. Ko also acknowledges the support of the Eric and Wendy Schmidt AI in Science Postdoctoral Fellowship, a Schmidt Futures program. We also acknowledged AdvanceSoft Corporation for implementing the LAMMPS interface.

\end{acknowledgement}



\section*{Supplementary material}

# SUPPLEMENTARY INFORMATION

# Materials Graph Library (MatGL), an open-source graph deep learning library for materials science and chemistry

Tsz Wai Ko,*,† Bowen Deng,‡,¶ Marcel Nassar,§ Luis Barroso-Luque,‡,¶ Runze Liu,† Ji Qi,† Elliott Liu,† Gerbrand, Ceder,‡,¶ Santiago Miret,§ and Shyue Ping Ong*,†

†*Aiiso Yufeng Li Family Department of Chemical and Nano Engineering, University of California San Diego, 9500 Gilman Dr, Mail Code 0448, La Jolla, CA 92093-0448, United States*

‡*Department of Materials Science and Engineering, University of California Berkeley, Berkeley, CA, USA*

¶*Materials Sciences Division, Lawrence Berkeley National Laboratory, California 94720, United States*

§*Intel Labs, Santa Clara, CA, United States*

E-mail: t1ko@ucsd.edu; ongsp@ucsd.edu

# Learning Curves of Property Models

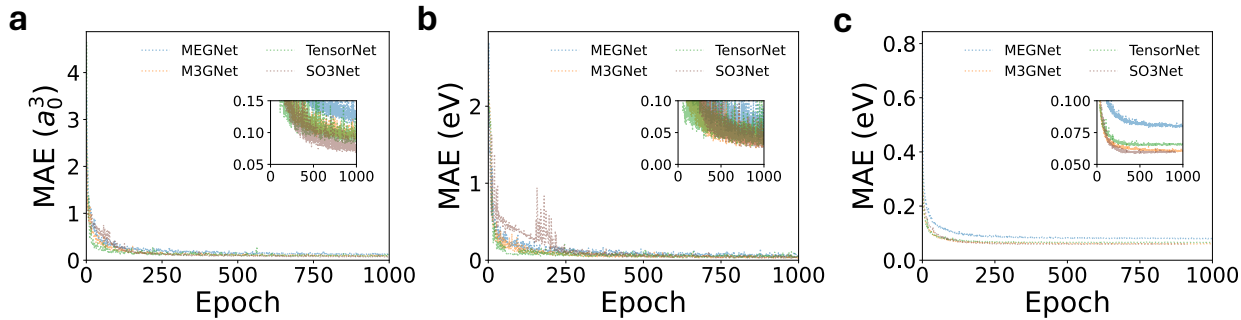

Fig. S1: **Learning curves of property models for QM9.** The training and validation mean absolute errors (MAE) of **a** isotropic polarizability $\alpha$, **b** free energy $G$ and **c** gap between LUMO and HOMO $\Delta\epsilon$ are computed with various graph neural networks including MEGNet, M3GNet, TensorNet and SO3Net during the training. The small figure shows a closer look into the validation error of different models for better visualization. The target properties are token from the QM9 benchmark dataset.

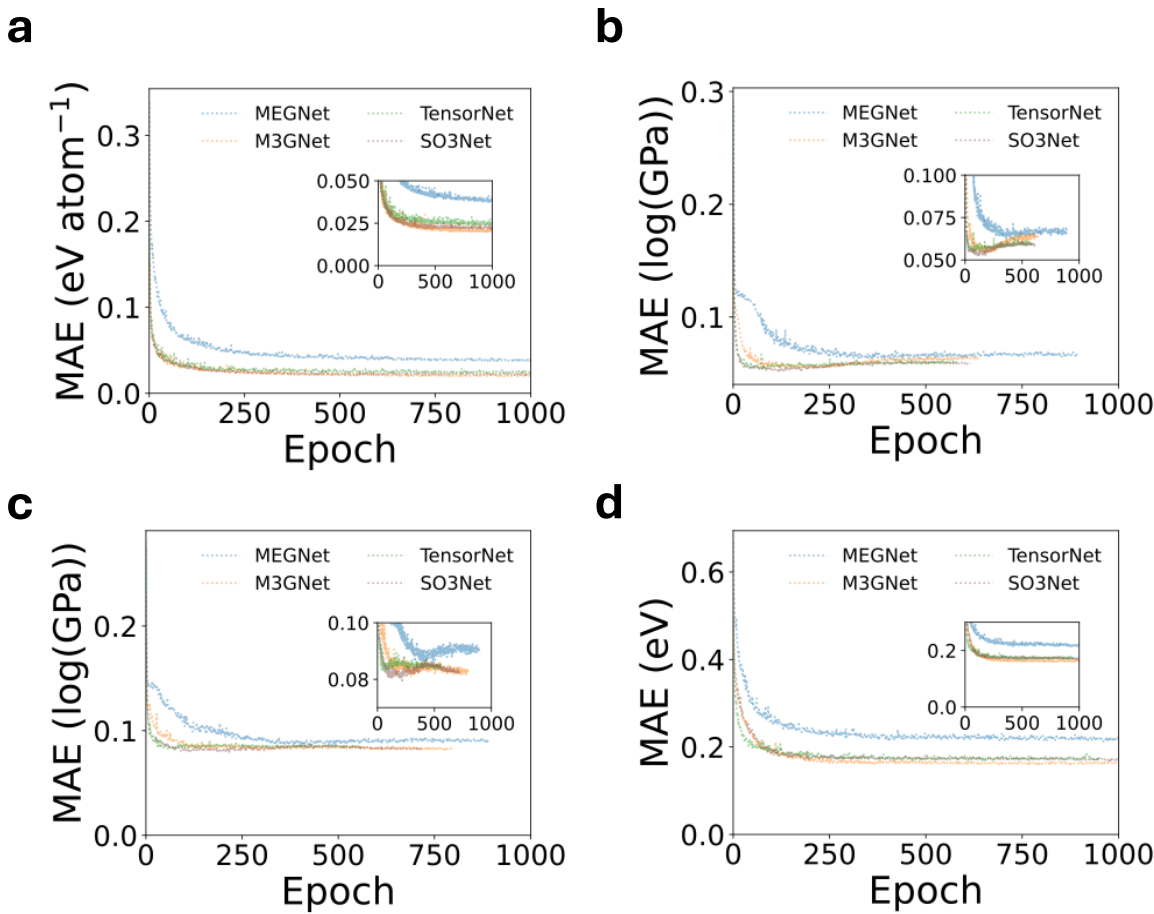

Fig. S2: **Learning curves of property models for Matbench.** The training and validation mean absolute errors (MAE) of **a** formation energy $E_{\text{form}}$, **b** Voigt-Reuss-Hill bulk modulus $\log(K_{\text{VRH}})$, **c** shear modulus $\log(G_{\text{VRH}})$ and **d** bandgap ($E_{\text{G}}$) are computed with various graph neural networks including MEGNet, M3GNet, TensorNet and SO3Net during the training. The small figure shows a closer look into the validation error of different models for better visualization. The target properties are token from the Matbench benchmark dataset.

3

# Learning Curves of Machine Learning Interatomic Potentials

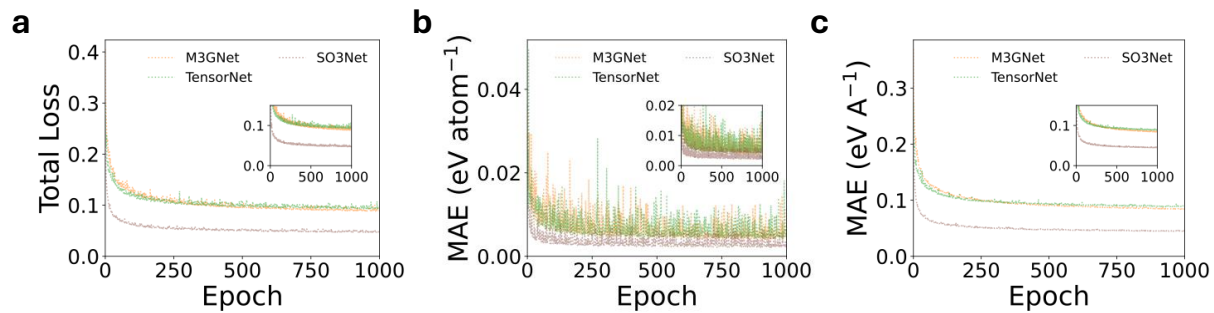

Fig. S3: **Learning curves of machine learning interatomic potentials for ANI-1x Subset.** The convergence of **a** Total validation loss, mean absolute error of **b** total energy and **c** force for M3GNet, TensorNet and SO3Net during the training.

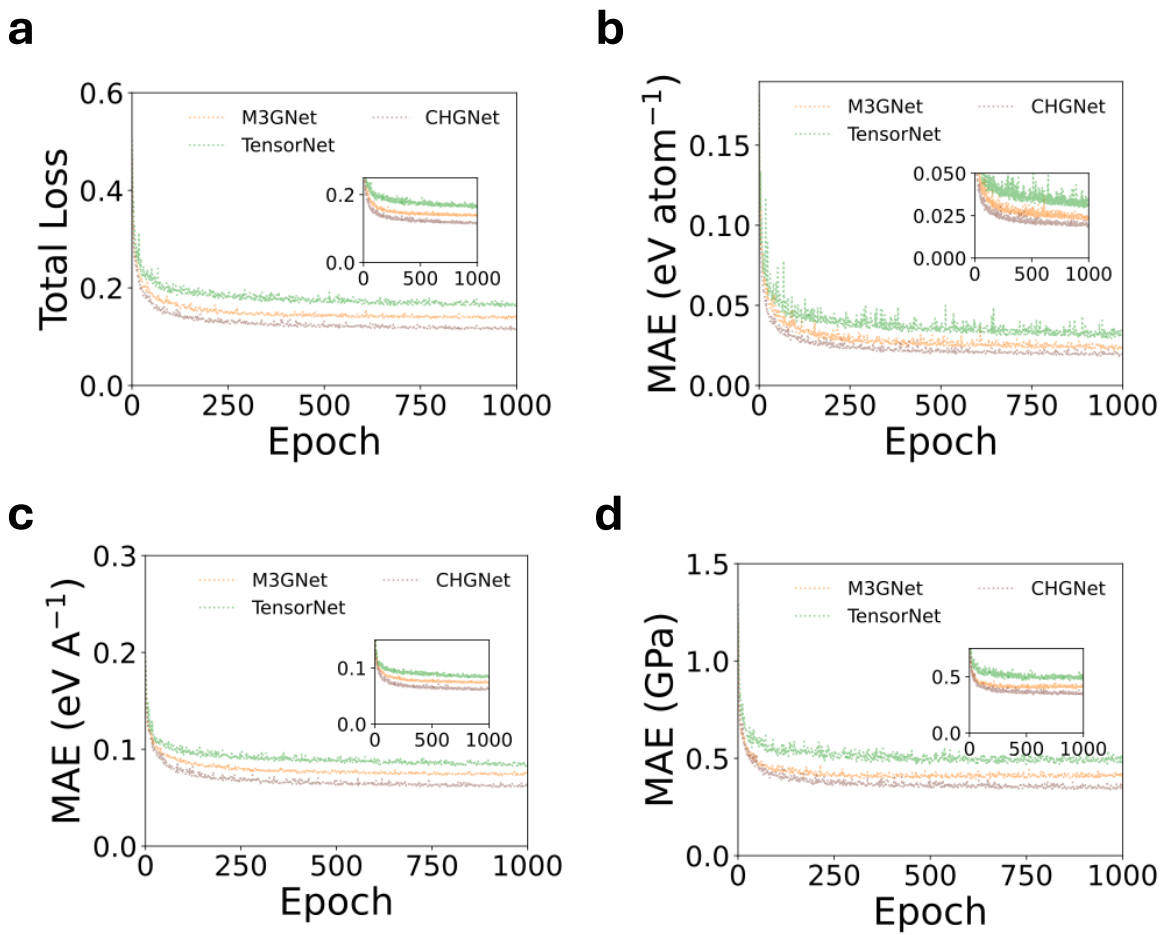

Fig. S4: **Learning curves of machine learning interatomic potentials for MPF-2021.2.8 Subset.** The convergence of **a** Total validation loss, mean absolute error of **b** total energy, **c** force and **d** stress for M3GNet, TensorNet and CHGNet during the training.

5

# Hyperparameters for Data-piplines

Table S1: **Input settings of MGLDataset and MGLDataLoader for property models and machine learning interatomic potentials.**

```
{
    "cutoff": 5.0
    "converter" : Structure2Graph/Molecule2Graph,
    "include_line_graph" : True/False,
    "include_directed_graph" : True/False,
    "threebody_cutoff" : None/3.0/4.0,
    "structures": list[Structure/Molecule],
    "labels": list[target_properties],
    "raw_dir": "./",
    "save_dir": "./"
}
```

Table S2: **Input settings for MGLDataLoader.**

```
{
    "train_data": training_set,
    "val_data" : validation_set,
    "test_data": test_set,
    "collate_fn": collate_fn_graph/collate_fn_pes,
    "batch_size": 32/64/128,
    "num_workers": 0,
}
```

# Input Parameters for Graph Deep Learning Models

The following tables summarize the input settings for different graph deep learning model architectures used to train intensive property models and extensive machine learning interatomic potentials.

Table S3: **Input settings for MEGNet.**

```
{
        "@class": "MEGNet",
        "@module": "matgl.models._megnet",
        "@model_version": 1,
        "init_args": {
            "dim_node_embedding": 16,
            "dim_edge_embedding": 100,
            "dim_state_embedding": 2,
            "ntypes_state": null,
            "nblocks": 3,
            "hidden_layer_sizes_conv": [64, 64, 32],
            "hidden_layer_sizes_output": [32, 16],
            "nlayers_set2set": 3,
            "niters_set2set": 3,
            "activation_type": "softplus2",
            "is_classification": False,
            "include_state": True,
            "dropout": 0.0,
            "element_types": DEFAULT_ELEMENTS
            "bond_expansion": null,
            "cutoff": 5.0,
            "gauss_width": 0.5,
            "hidden_layer_sizes_input": [64, 32],
            "is_intensive": True,
            "readout_type": "set2set"
        }
}
```

Table S4: **Input settings for M3GNet.**

```json
{
    "@class": "M3GNet",
    "@module": "matgl.models._m3gnet",
    "@model_version": 1,
    "init_args": {
        "element_types": DEFAULT_ELEMENTS,
        "dim_state_embedding": 0,
        "ntypes_state": null,
        "max_n": 3,
        "max_l": 3,
        "nblocks": 3,
        "rbf_type": "SphericalBessel",
        "is_intensive": True/False,
        "readout_type": "set2set",
        "task_type": "regression",
        "cutoff": 5.0,
        "threebody_cutoff": 4.0,
        "ntargets": 1,
        "use_smooth": True,
        "use_phi": False,
        "niters_set2set": 3,
        "nlayers_set2set": 3,
        "field": "node_feat",
        "activation_type": "swish",
        "dim_edge_embedding": 64,
        "dim_node_embedding": 64,
        "dim_state_feats": null,
        "include_state": False,
        "units": 64
    }
}
```

Table S5: **Input settings for TensorNet.**

```
{
    "@class": "TensorNet",
    "@module": "matgl.models._tensornet",
    "@model_version": 1,
    "init_args": {
        "element_types": DEFAULT_ELEMENTS,
        "ntypes_state": null,
        "dim_state_embedding": 0,
        "dim_state_feats": null,
        "include_state": False,
        "max_n": 3,
        "max_l": 3,
        "rbf_type": "SphericalBessel",
        "use_smooth": True,
        "activation_type": "swish",
        "width": 0.5,
        "readout_type": "set2set",
        "task_type": "regression",
        "niters_set2set": 3,
        "nlayers_set2set": 3,
        "field": "node_feat",
        "is_intensive": True/False,
        "ntargets": 1,
        "cutoff": 5.0,
        "dtype": "torch.float32",
        "equivariance_invariance_group": "O(3)",
        "nblocks": 2,
        "num_rbf": 32,
        "units": 64
    }
}
```

Table S6: **Input settings for SO3Net**

```
{
    "@class": "SO3Net",
    "@module": "matgl.models._so3net",
    "@model_version": 0,
    "init_args": {
        "element_types": DEFAULT_ELEMENTS,
        "units": 64,
        "dim_state_embedding": 0,
        "ntypes_state": null,
        "dim_state_feats": null,
        "nblocks": 3,
        "nmax": 5,
        "cutoff": 5.0,
        "rbf_learnable": False,
        "target_property": "graph",
        "task_type": "regression",
        "readout_type": "set2set",
        "niters_set2set": 3,
        "nlayers_set2set": 3,
        "nlayers_readout": 2,
        "is_intensive": True,
        "include_state": False,
        "use_vector_representation": False,
        "correct_charges": False,
        "predict_dipole_magnitude": False,
        "activation_type": "swish",
        "ntargets": 1,
        "return_vector_representation": False,
        "dim_node_embedding": 64,
        "lmax": 2
    }
}
```

Table S7: **Input settings for CHGNet.**

```
{
    "@class": "CHGNet",
    "@module": "matgl.models._chgnet",
    "@model_version": 1,
    "init_args": {
        "element_types": DEFAULT_ELEMENTS,
        "dim_state_feats": null,
        "non_linear_bond_embedding": False,
        "non_linear_angle_embedding": False,
        "cutoff": 5.0,
        "threebody_cutoff": 3.0,
        "cutoff_exponent": 3,
        "max_f": 3,
        "learn_basis": False,
        "num_blocks": 3,
        "shared_bond_weights": "both",
        "final_mlp_type": "mlp",
        "final_hidden_dims": [64, 64, 64],
        "final_dropout": 0.0,
        "pooling_operation": "sum",
        "readout_field": "atom_feat",
        "activation_type": "swish",
        "is_intensive": False,
        "num_targets": 1,
        "num_site_targets": 1,
        "task_type": "regression",
        "angle_update_hidden_dims": [],
        "atom_conv_hidden_dims": [64],
        "bond_conv_hidden_dims": [64],
        "bond_update_hidden_dims": null,
        "conv_dropout": 0.0,
        "dim_angle_embedding": 64,
        "dim_atom_embedding": 64,
        "dim_bond_embedding": 64,
        "dim_state_embedding": null,
        "layer_bond_weights": null,
        "max_n": 3,
        "normalization": null,
        "normalize_hidden": False
    }
}
```



\end{document}